\documentclass[aps,pra,twocolumn,showpacs,superscriptaddress]{revtex4} 
\usepackage{graphicx}
\usepackage{amsmath}
\usepackage{amssymb}
\usepackage{epstopdf}
\usepackage{amsfonts}
\usepackage{dcolumn}
\begin{document}
\title{Harmonically Trapped Four-Boson System}
\author{D. Blume}
\affiliation{Homer L. Dodge Department of Physics and Astronomy,
  The University of Oklahoma,
  440 W. Brooks Street,
  Norman,
  Oklahoma 73019, USA}
\author{M. W. C. Sze}
\affiliation{JILA, NIST, and Department of Physics, University of Colorado,
  Boulder, Colorado 80309-0440, USA}
\author{J. L. Bohn}
\affiliation{JILA, NIST, and Department of Physics, University of Colorado,
  Boulder, Colorado 80309-0440, USA}
\date{\today}

\begin{abstract}
Four identical spinless bosons with purely attractive
two-body short-range interactions and repulsive three-body
interactions 
under external spherically symmetric harmonic confinement
are considered.
The repulsive three-body potential prevents 
the formation of deeply-bound states with molecular character.
The low-energy spectrum with
vanishing orbital
angular momentum and positive parity for infinitely large two-body
$s$-wave scattering length is analyzed in detail.
Using the three-body contact,
states are classified as universal, quasi-universal,
or strongly non-universal.
Connections with the zero-range interaction model
are discussed.
The energy spectrum is mapped out as
a function of the two-body $s$-wave scattering length $a_s$, $a_s>0$.
In the weakly- to medium-strongly-interacting regime, one of the states
approaches the energy obtained for a hard core interaction model.
This state is identified as the energetically lowest-lying ``BEC state''.
Structural properties are also presented.
\end{abstract}
\maketitle

\section{Introduction}
\label{introduction}
The unitary two-component Fermi gas is realized when the 
$s$-wave scattering length $a_s$ between the spin-up
and spin-down atoms is infinitely
large~\cite{giorginiRMP}.
In the limit that the range of the interspecies interactions goes
to zero, the unitary Fermi gas has been shown to be fully universal,
i.e., the infinitely strongly-interacting gas is
characterized by the same number of length
scales as the non-interacting
gas~\cite{baker1999,cowell2002,werner2006PRA}.
The relevant length
scales in the universal
regime are the de Broglie wave length and the average interparticle
spacing that is related to the density of the 
homogeneous system~\cite{ho2004}. For the harmonically
confined system with angular frequency $\omega$,
the latter scale is typically replaced by the harmonic oscillator length
$a_{\text{ho}}$, where $a_{\text{ho}}=\sqrt{\hbar/(m \omega)}$. Here,
$m$ denotes the atom mass.

Universality implies that the realistic two-body interaction
potentials can be replaced by a two-body boundary condition on the
many-body wave function $\Psi_{\text{tot}}$ in the limit that
the interparticle distance
coordinate $r_{jk}$ of particle $j$ 
(a spin-up atom) and particle $k$ (a spin-down atom)
goes to zero
while all other coordinates
$(\vec{r}_j+\vec{r}_k)/2,\vec{r}_1,\cdots,\vec{r}_{j-1},\vec{r}_{j+1},
\cdots,\vec{r}_{k-1},\vec{r}_{k+1},\cdots,\vec{r}_N$
are being held fixed:
\begin{eqnarray}
\label{eq_bc}
\lim_{r_{jk} \rightarrow 0}
\frac{1}{r_{jk} \Psi_{\text{tot}}}
\frac{\partial (r_{jk} \Psi_{\text{tot}})}{\partial r_{jk}}
= -\frac{1}{a_s}.
\end{eqnarray} 
Here, $N$ denotes the total number of particles and 
$\vec{r}_j$ the position vector of the $j$-th particle.
No boundary conditions in the three- or higher-body
sectors are needed, i.e., the Pauli exclusion principle ``naturally''
guarantees that the probability to find three or more
particles on top of each other vanishes.
The infinitely strongly-interacting 
two-component Fermi gas is found to be mechanically stable and three-body 
losses are so low that the lifetime is large compared
to the time scale associated with the Fermi energy and 
the other time scales of the system~\cite{regal2004,zwierlein2004}.

The situation for identical bosons is fundamentally different
than that for two-component fermions.
Unless the two-body boundary condition [see Eq.~(\ref{eq_bc})],
which is enforced for every pair, is supplemented
by a three-body boundary condition
(this can be achieved through a three-body potential or
via a momentum cut-off, among others), 
the lowest energy of the Bose gas at unitarity is 
unbounded from below, i.e., the Bose gas undergoes a generalization
of the Thomas collapse, which was first
studied for three particles by Thomas in 1935 in the context of the
triton~\cite{thomas1935}.
The Thomas collapse is intimately related to the
existence of an infinite tower of
Efimov states for three identical bosons in free
space~\cite{efimov70,braaten2006,naidon2017}
(no external confining potential).

When approaching the low temperature unitary regime adiabatically,
either by tuning the $s$-wave scattering length at low temperature
or by decreasing the temperature at large $s$-wave scattering 
length~\cite{navon2011,wild2012,rem2013,fletcher2013,makotyn2014,eismann2016,klauss2017,fletcher2017},
the system exhibits detrimental losses
due to three-body recombination by which
energetic atoms are being expelled from the trap.
These three-body recombination processes are governed by
Efimov physics~\cite{esry1999,nielsen1999,bedaque2000,grimm2006}.
Probing the gas at unitarity therefore requires jumping rapidly
and non-adiabatically to this regime.
Following non-adiabatic pathways,
the unitary Bose gas---presumably in local but not
in global equilibrium---has been probed experimentally
using interferometric
protocols~\cite{fletcher2013,klauss2017}.
The time-dependent contact at unitarity has been extracted and evidence
that three-body Efimov states
are being occupied during the non-adiabatic
ramp sequence has been presented~\cite{wild2012,klauss2017,fletcher2017}.

The theoretical treatment of the unitary Bose gas is highly
non-trivial and a variety of approximate
static and dynamic techniques
have been
applied~\cite{song2009,lee2010,yin2013,zhou2013,sykes2014,corson2015,ancilotto2015,yin2016,jiang2016,ding2017,colussi2017},
producing results that do not seem
to yield
a simple consistent physical picture.
One line of work considers two- or three-body systems 
with the external confinement tightened such that the density
of the few-body system is comparable to that realized experimentally
for $N \approx 10^5$ atoms~\cite{klauss2017,sykes2014,colussi2017}.
Motivated by the physical insights that have already been gained from
these two- and three-body studies,
the present work considers the next larger system, namely
the harmonically trapped four-body system,
for various positive two-body $s$-wave scattering lengths
ranging from zero to infinity.
The numerically obtained energy spectrum and the associated 
eigen states are characterized.

The remainder of this paper is organized as follows.
Section~\ref{sec_background} introduces the
system Hamiltonian and pertinent theoretical background.
Sections~\ref{sec_n3} and \ref{sec_n4}
present our results for the harmonically trapped
three- and four-boson systems, respectively.
Lastly, Sec.~\ref{sec_conclusion}
presents a summary and an outlook.

\section{System Hamiltonian}
\label{sec_background}

We consider $N$ identical bosons with mass $m$ under
spherically symmetric external confinement
with angular frequency $\omega$.
The system Hamiltonian $H_{\text{tot}}$,
\begin{eqnarray}
H_{\text{tot}}=H_{\text{cm}} + H,
\end{eqnarray}
can be divided into the center-of-mass Hamiltonian $H_{\text{cm}}$,
\begin{eqnarray}
H_{\text{cm}} = \frac{-\hbar^2}{2M} \nabla_{\vec{R}_{\text{cm}}}^2 + \frac{1}{2}
M \omega^2 \vec{R}_{\text{cm}}^2,
\end{eqnarray}
and the relative Hamiltonian $H$,
\begin{eqnarray}
H = \sum_{j=1}^{N-1} \left( \frac{-\hbar^2}{2 \mu_j} \nabla_{\vec{\rho}_j}^2
+ 
\frac{1}{2} \mu_j \omega^2 \vec{\rho}_j^2 \right)
+ V_{\text{int}}(\vec{\rho}_1,\cdots,\vec{\rho}_{N-1}).
\end{eqnarray}
Here, $M$ denotes the total mass, $M=Nm$, and $\vec{R}_{\text{cm}}$
the center-of-mass vector,
$\vec{R}_{\text{cm}} = N^{-1} (\vec{r}_1+\cdots+\vec{r}_N)$,
with $\vec{r}_j$ denoting the position
vector of the $j$-th particle measured with respect 
to the center of the trap.
The relative Hamiltonian $H$ is written in terms of the
Jacobi vectors $\vec{\rho}_j$ and the associated 
Jacobi masses $\mu_j$.
For the purpose of this work, the 
explicit definition of the Jacobi coordinates is
irrelevant.
The important point is that the interaction potential
$V_{\text{int}}$
is independent of the center-of-mass vector $\vec{R}_{\text{cm}}$.
Correspondingly, the eigen states $\Psi_{\text{tot}}$ of $H_{\text{tot}}$
separate into a center-of-mass piece $\Psi_{\text{cm}}(\vec{R}_{\text{cm}})$
and a relative piece
$\Psi(\vec{\rho}_1,\cdots,\vec{\rho}_{N-1})$,
\begin{eqnarray}
\Psi_{\text{tot}}(\vec{r}_1,\cdots,\vec{r}_N)=
\Psi_{\text{cm}}(\vec{R}_{\text{cm}}) \Psi(\vec{\rho}_1,\cdots,\vec{\rho}_{N-1}).
\end{eqnarray}
Since the Schr\"odinger equation for the center-of-mass vector $\vec{R}_{\text{cm}}$
is identical to that of a three-dimensional harmonic oscillator, the
solutions can be written down readily.
Our goal in this paper is to solve the relative Schr\"odinger equation
\begin{eqnarray}
H \Psi(\vec{\rho}_1,\cdots,\vec{\rho}_{N-1}) = 
E \Psi(\vec{\rho}_1,\cdots,\vec{\rho}_{N-1}).
\end{eqnarray}

Realistic atom-atom interaction potentials support
many two-body bound states. 
Consequently,
the degenerate $N$-body gas 
(even the two-component Fermi gas) corresponds to a 
highly-excited metastable state.
To make the $N$-body problem tractable, we work with
low-energy interactions that eliminate, from the outset,
 $N$-body states that
correspond to deeply-bound molecular clusters.
It should be noted that this simplification also eliminates
decay channels that are needed for three-body recombination processes,
which are known to play
a role in the quench experiments, to occur.
In these experiments, 
the Bose gas disappears before global equilibrium is reached,
restricting observations to systems that are in local equilibrium but
not in global equilibrium.
This work considers three different low-energy interactions.

\subsection{Interaction Model I}
Model~I assumes two-body hard core interactions,
\begin{eqnarray}
V_{\text{int}}^{\text{I}} = \sum_{j=1}^{N-1} \sum_{k>j}^N V_{\text{2b}}^{\text{hc}}(r_{jk}),
\end{eqnarray}
where
\begin{eqnarray}
V_{\text{2b}}^{\text{hc}}(r_{jk}) =  \Big \{ \begin{array}{lll}
\infty & \mbox{for} & r_{jk} \le a_s \\
0 & \mbox{for} & r_{jk} > a_s \end{array}
.
\end{eqnarray} 
This model interaction, which has been used extensively
in the literature (see, e.g., 
Refs.~\cite{kalos1974,giorgini1999,blume2001}),
is expected to provide a reliable description
of the weakly-repulsive Bose gas.
Since the range of the hard core potential increases with
increasing $a_s$, this interaction yields model-dependent
results when $a_s/a_{\text{ho}}$ is not small. 
The ground state of the time-independent 
$N$-particle Schr\"odinger equation 
for this model interaction can be found efficiently
using the diffusion quantum Monte Carlo
technique~\cite{mcbook,blume2001}.
For positive and sufficiently small $a_s$ (see Secs.~\ref{sec_n3}
and \ref{sec_n4} for more quantitative
statements),
this ground state corresponds to
the gas-like state that we are interested in.
No excited states are considered for model~I.

\subsection{Interaction Model II}
Model~II assumes an attractive two-body Gaussian 
interaction
$V_{\text{2b}}^{\text{G}}(r_{jk})$ with depth $v_0$ ($v_0 < 0$) and
range $r_0$,
\begin{eqnarray}
V_{\text{2b}}^{\text{G}}(r_{jk}) = v_0 \exp \left[ - \left(
\frac{r_{jk}}{\sqrt{2} r_0} \right) ^2
\right].
\end{eqnarray}
Throughout this 
work,
the range is fixed at $r_0=0.025 a_{\text{ho}}$ and the depth $v_0$
is adjusted to dial in the desired
$s$-wave scattering length $a_s$
($a_s \ge 0$).
We only consider depths for which
the Gaussian potential supports
one two-body $s$-wave bound state in
free space (at unitarity, this
bound state has zero energy).
To prevent the formation of deeply-bound molecular 
three- and four-body states,
a repulsive three-body Gaussian potential
$V_{\text{3b}}^{\text{G}}(r_{jkl})$ with height $V_0$ and range $R_0$ is added,
\begin{eqnarray}
V_{\text{3b}}^{\text{G}}(r_{jkl}) = V_0 \exp \left[ - \left(
\frac{r_{jkl}}{\sqrt{2} R_0} \right) ^2 \right]
.
\end{eqnarray}
Here, $r_{jkl}$ denotes the ``triple sub-hyperradius'',
\begin{eqnarray}
r_{jkl} = \sqrt{ r_{jk}^2 + r_{jl}^2 + r_{kl}^2 
}.
\end{eqnarray}
For the $N=3$ system, there exists one such triple sub-hyperradius, which
coincides
with the $N$-body hyperradius $R$,
\begin{eqnarray}
R =  \sqrt{ \sum_{j=1}^{N-1} \sum_{k>j}^N r_{jk}^2}.
\end{eqnarray}
For the $N=4$ system, there exist four triple sub-hyperradii.
Throughout, we fix $R_0$ at $\sqrt{8} r_0 \approx 0.071 a_{\text{ho}}$.
The height $V_0$ is varied to investigate the dependence of the
results on the three-body potential.
Putting things together, interaction model~II reads
\begin{eqnarray}
V_{\text{int}}^{\text{II}}=
\sum_{j=1}^{N-1} \sum_{k>j}^N V_{\text{2b}}^{\text{G}}(r_{jk})+
\sum_{j=1}^{N-2} \sum_{k>j}^{N-1} \sum_{l>k}^N 
V_{\text{3b}}^{\text{G}}(r_{jkl}).
\end{eqnarray}

While the repulsive 
three-body Gaussian potential eliminates deep-lying molecular states,
the resulting Hamiltonian supports states with gas-like and 
with molecular-like
character (see Secs.~\ref{sec_n3} and \ref{sec_n4} for
details).
To obtain the entire low-energy spectrum and to
subsequently analyze  what the characteristics
of the eigen states are, we employ a basis set expansion approach.
The eigen states $\Psi$ are expanded in terms of a set
of non-orthogonal explicitly correlated Gaussian basis functions,
which depend on the interparticle distance
coordinates and a set of non-linear variational   
parameters~~\cite{cgbook,mitroy2013}.
The non-linear variational parameters, which are optimized
semi-stochastically, give the flexibility
to describe correlations at different length scales,
namely, at the scales of the range
of the interaction potential and 
the harmonic oscillator length.
The
resulting generalized eigenvalue
problem is solved through diagonalization~\cite{cgbook,mitroy2013}.
The resulting eigen values are, according to
Ritz' variational principle, variational upper bounds to the true
eigen energies (this statement holds for the ground and the excited
states)~\cite{cgbook}.
Throughout this paper, only states $\Psi$ with
vanishing relative orbital angular momentum $L$ and positive parity 
$\Pi$ are considered.
These states constitute a small subset of the entire set of eigen states.

As the two-body depth $v_0$ becomes more negative, the 
four-body system supports an increasing number of states with negative 
energy. As a consequence, the lowest BEC state 
(see below for a definition) is a fairly
highly-excited state of the Hamiltonian with interaction model~II.
For $v_0$ not too negative, we optimize one eigen state
at a time. We follow the same general strategy
as the one that was pursued to determine
a
large number of eigen energies and eigen states 
of the two-component
four-particle Fermi gas~\cite{rakshit2012}. 
The basis set 
that describes, e.g., the tenth eigen state accurately
may describe the nine energetically lower-lying
states comparatively poorly. However, as long as the
basis captures the nine lower-lying states, even if poorly,
the energy of the tenth state provides an upper variational bound for
the tenth state. 
A typical basis set
consists of 600 (for low state numbers) to 2,000
(for large state numbers) fully symmetrized basis functions.

The portion of the four-body spectrum that we are 
interested in consists of more and more highly excited states as the
$s$-wave scattering length decreases (i.e., as $v_0$ becomes more 
negative). 
As a consequence, it becomes increasingly tedious to 
generate a basis set for each eigen state. In particular, even if we do not
optimize each state, to obtain a tight bound
for the energy of, say, state number 70, we still need to
describe 69 lower-lying states.
As an alternative, we 
employ a ``target  state approach'', which optimizes the state
whose energy is greater than but closest to a preset 
target energy. 
The target state approach is similar in spirit
to what has been used to identify resonance 
states (see, e.g., Ref.~\cite{usukura2002}).
For example, if we expect that the system supports a
state with energy $E^*$, then we set the target energy
to $E^*-\Delta E$, where $\Delta E$ is positive and of the order
of a tenth of $E_{\text{ho}}$. When we enlarge the basis set, we 
add basis functions that lower the energy of the state whose energy
is above the target energy and closest to it. 
As more basis functions are added, the
energy of this state drops below $E^*-\Delta E$ and we optimize the
state with the next larger energy. When plotting the eigen energies
as a function of the inverse of the number of basis functions,
we observe convergence of the energy corresponding to different
state numbers. 
If we repeat the calculation for different $\Delta E$ and find the same
final energy, then we can be fairly sure that we have found an
isolated eigen energy, i.e., the energy of an eigen state
away from avoided crossings.
Since we do not, in this case, describe all the lower-lying states,
the resulting energy, extrapolated  to the infinite basis set limit,
does not provide a strict variational upper bound.
The advantage of this approach is that the majority of the basis
functions added is used to improve the description of the
state of interest without having to describe (even if relatively poorly)
all the lower-lying states. 

\subsection{Interaction Model III}
Model~III employs two-body zero-range interactions.
As already alluded  to in the introduction, the two-body interactions can be
accounted for by
enforcing the boundary condition in Eq.~(\ref{eq_bc})
on the relative many-body eigen state $\Psi$.
Much of the remainder of this subsection focuses on unitarity.

At unitarity, the non-interacting Hamiltonian with the
boundary condition given in Eq.~(\ref{eq_bc})
supports relative eigen states that can be written as
a direct product
of a function $\Phi_{\nu q}$ that depends on the 
four-body hyperradius $R$
and a function $\phi_{\nu}$ that is independent 
of $R$~\cite{werner2006PRA,castin2004,tan2004},
\begin{eqnarray}
\label{eq_wavefct_unit_universal}
\Psi_{\nu q}(\vec{\rho}_1,\cdots,\vec{\rho}_{N-1})=
R^{-(3N-4)/2}
\Phi_{\nu q}(R) \phi_{\nu}(\vec{\Omega}).
\end{eqnarray}
Here, $\vec{\Omega}$ collectively denotes
the $3N-4$ hyperangular coordinates
(the definition of the hyperangles is not important for the 
discussion that follows).
The eigen energies corresponding to these eigen states
are given by
\begin{eqnarray}
\label{eq_energy_unit_universal}
E_{\nu q}^{\text{unit}} = 
( 2 q + s_{\nu} + 1) E_{\text{ho}},
\end{eqnarray}
where $q = 0, 1,\cdots$ denotes the
hyperradial quantum number and
$s_{\nu}$ (which is real-valued) 
would be obtained by solving a differential equation
in the hyperangular degrees of freedom.
The harmonic oscillator energy $E_{\text{ho}}$ is defined
through $E_{\text{ho}}= \hbar \omega$.
The quantum number $\nu$ enumerates the solutions of the 
differential equation in the hyperangular coordinates.
Equation~(\ref{eq_energy_unit_universal}) is a consequence of the
fact that each hyperradial potential curve, which is characterized
by $s_{\nu}$, supports a ladder of states with energy spacing
$2 E_{\text{ho}}$; this spacing 
is the same as for the non-interacting system
(the $s_{\nu}$ values are,
however, in general different)~\cite{werner2006PRA}.
The values of $s_{\nu}$ for the unitary Bose gas
are, for $N>3$, challenging to determine
and, to the best of our knowledge, not known.
For $N=3$, the $s_{\nu}$ values can be determined semi-analytically
with arbitrary accuracy~\cite{braaten2006,efimov1971,efimov1973}.
We refer to the states of the form given in 
Eq.~(\ref{eq_wavefct_unit_universal}) with eigen energies of the form
given in Eq.~(\ref{eq_energy_unit_universal})
as universal states.
Universal in this sense implies that the eigen energies
and eigen states are indifferent to the three-body potential, and 
are thus characterized by the two-body $s$-wave scattering
length alone.

Importantly, model~III
supports a second class of states at unitarity
that we refer to as non-universal
(for $N=3$, Ref.~\cite{wernerPRL2006} 
refers to this class of states as 
Efimovian).
To introduce these non-universal states, we consider the $N=3$ system.
The hyperangular equation for $N=3$, which is solved subject to the
two-body boundary condition, yields one imaginary
eigen value, namely $s_0 = \imath 1.00624\dots$.
Neglecting for the moment
the external confinement, this eigen value gives rise to a
purely attractive effective hyperradial
potential curve that is proportional to
$-(|s_0|^2+1/4)R^{-2}$ and supports an infinite number of three-body
bound states,
with the ratio between neighboring energy levels being 
equal to $\exp(\pi/|s_0|)^{-2}\approx 1/515$.
To fix the absolute position of the energy levels, a small-$R$
boundary condition on the 
hyperradial function $\Phi_0$ needs to be 
specified~\cite{efimov70,braaten2006,naidon2017}.
Such an additional boundary condition is not needed 
for the universal states, since
the probability to find three bosons at the same location is vanishingly 
small ``on its own'' due to the repulsive small-$R$ behavior of the effective
hyperradial potential curves
for real $s_{\nu}$ (if an additional small-$R$
boundary condition was added, the universal states
would be insensitive to it).
If the trap is added, the spacing between the energy of
consecutive non-universal states is modified drastically. 
The resulting energy spacings 
depend on $q$ and differ, in general,
from the $2 E_{\text{ho}}$ spacing that is
a key signature of
the universal 
states~\cite{jonsell2002,wernerPRL2006}.

The non-universal states of the trapped $N=4$ unitary system
have not yet been studied in much depth~\cite{toelle2011,toelle2013}.
When solving the hyperangular equation for $N=4$,
the three-body boundary condition for each triple should be accounted
for in addition
to the two-body boundary condition for each pair.
While this hyperangular equation has not yet been solved for
zero-range interactions,
it has been solved for finite-range interactions~\cite{vonstecher2009}.
The fixed-$R$ hyperangular eigenvalues were shown to depend on
the four-body hyperradius $R$, implying non-separability
of the hyperradial and hyperangular degrees of freedom.
Setting $\omega$ to zero,
there exist as many 
effective hyperradial potential curves as there exist 
three-body Efimov states and the large-$R$ values 
($R$ can be very large) of these potential curves
are just the 
energies of the three-body 
Efimov states~\cite{vonstecher2009}.
This is distinctly different from the three-boson system, which
supports one effective hyperradial potential 
curve at unitarity in which an infinity of
non-universal three-body states live
(the other potential curves support universal states).

The confining potential ``cuts off'' the asymptotic region
for all but the few 
lowest effective free-space hyperradial four-body
potential curves
(depending on the
system parameters, it might be all but one).
The term ``cuts off'' is used here to indicate that the asymptotic
behavior of the effective potential curves is suppressed by
the quadratically increasing confining potential, which forces the
wave function to fall off exponentially at much smaller $R$
than it would in the absence of the trap, where there
exist two four-body states that are tied to each Efimov 
trimer~\cite{vonstecher2009,platter2004}.
To determine the eigen energies of the non-universal states,
the coupling (which may be small)
between different potential curves
needs to be accounted for even at unitarity;
this is, again, distinctly different from the three-boson
system where the coupling at unitarity vanishes 
for the universal {\em{and}} the non-universal states.
It is expected that the resulting energy ladders 
for the non-universal states of the
trapped four-boson system 
are characterized by energy spacings that,
in general, differ from
$2 E_{\text{ho}}$.

The four-body results 
at unitarity presented in Sec.~\ref{sec_n4}
are not obtained by first solving the hyperangular differential
equation and by then subsequently solving a set of coupled differential 
equations in the hyperradius. Instead, the four-body results
are obtained by treating all relative
degrees of freedom on equal footing.
The resulting energy spectrum and eigen functions
are, however, analyzed within the
hyperspherical coordinate framework introduced above.
A main outcome of the analysis for
$N=4$ will be that some
states do not
fit neatly into the categories of universal or
non-universal. 
We will denote these states as quasi-universal.
To illustrate aspects of our approach, the next section
presents three-body results for interaction models~I, II, and III.

Moving from unitarity to finite $s$-wave scattering lengths,
the eigen states 
of the harmonically trapped $N$-boson system cannot
{\em{a priori}} be
divided into
universal and non-universal states even in principle.
Separability between the hyperangular and hyperradial degrees of freedom
does generally not exist and the energy spectrum is expected to exhibit
series of avoided crossings.
The next section illustrates that one can, away from the
avoided crossings, nevertheless meaningfully categorize states
as universal and non-universal.

\section{$N=3$ System}
\label{sec_n3}
\begin{figure}[t]
  \vspace*{0.6in}
  \hspace*{0.1in}
\centering
\includegraphics[angle=0,width=0.4\textwidth]{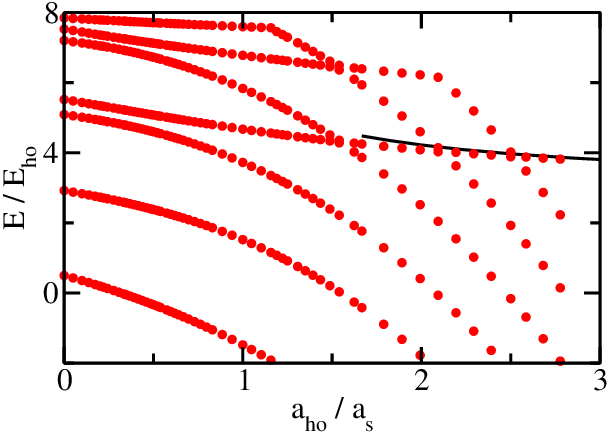}
\vspace*{.6in}
\caption{(Color online)
Energy spectrum for three harmonically
trapped identical bosons as a function of $a_{\text{ho}}/a_s$.
Circles show the relative energy of the seven energetically lowest-lying states 
with $L^{\Pi}=0^+$ symmetry 
for model~II with 
$V_0=97,700 E_{\text{ho}}$.
The solid line shows the relative energy of the lowest
state for two-body hard core interactions (model~I).
}\label{fig_energy_n3}
\end{figure}

To set the stage for our $N=4$ results,
this section summarizes selected results
for the $N=3$ system.
Circles in Fig.~\ref{fig_energy_n3}
show the relative three-body spectrum for the
interaction model~II as a function of the inverse of the 
$s$-wave scattering length for a fixed three-body interaction;
specifically, $V_0$ is set to $97,700 E_{\text{ho}}$.
At unitarity,
the 
$(\nu,q)$ quantum numbers (see Table~\ref{tab_n3}) 
are assigned as follows.
Using the two smallest real $s_{\nu}$ 
($s_{1}=4.46529\dots$ and $s_2=6.81836 \dots$)
for the zero-range model in Eq.~(\ref{eq_energy_unit_universal}),
the following relative zero-range energies at unitarity are found:
$E_{1,0}^{\text{unit}}= 5.46529 E_{\text{ho}}$,
$E_{1,1}^{\text{unit}}= 7.46529 E_{\text{ho}}$, and
$E_{2,0}^{\text{unit}}= 7.81836 E_{\text{ho}}$~\cite{wernerPRL2006}.
These zero-range energies for the universal states
agree well with a subset of our numerical energies for 
model~II:
$E=5.5187 E_{\text{ho}}$, $7.5258 E_{\text{ho}}$, and $7.8452 E_{\text{ho}}$.
The differences, which are of the order
of 1~\%, are attributed to the fact that the range
of our two-body Gaussian potential is $r_0=0.025 a_{\text{ho}}$ and not
zero.
The  hyperradial densities $P(R)$, obtained for model~II,
confirm this assignment.
For example, the hyperradial density for
a state with quantum number  $q$ has $q-1$ zeroes along the hyperradial
coordinate.

The remaining energies at unitarity (see Table~\ref{tab_n3})
are identified as belonging to
non-universal states, i.e., to states that are supported
by the effective hyperradial potential curve labeled by
$\nu=0$ (imaginary $s_0$).
The corresponding energy spacings  ($2.4217 E_{\text{ho}}$,
$2.1678 E_{\text{ho}}$,
$2.1110 E_{\text{ho}}$,
and $2.0869 E_{\text{ho}}$)
deviate notably from $2 E_{\text{ho}}$;
the deviations decrease
with increasing $q$.
In the absence of the harmonic trap, the lowest
relative three-body energy is 
$E_{\text{fs}}=-1.19 \times 10^{-4} \hbar^2/(m r_0^2) $. 
Using this to estimate the size $L_{\text{fs}}$ of the free-space trimer
at unitarity
via the expression $L_{\text{fs}}=(\kappa^*_{\text{fs}})^{-1}$,
where the binding momentum $\kappa_{\text{fs}}$ is defined through
$\kappa_{\text{fs}}=\sqrt{m |E_{\text{fs}}|/\hbar^2}$
and $\kappa_{\text{fs}}^*$ denotes the binding momentum at unitarity,
we find 
$L_{\text{fs}}=91.5r_0$ or, 
using $r_0=0.025a_{\text{ho}}$, $L_{\text{fs}}=2.29 a_{\text{ho}}$,
i.e., the lowest free-space Efimov trimer is roughly
of the same size as the harmonic oscillator
length of the external confinement.
If we use the lowest relative 
three-body energy for the finite-range
model~II of the trapped three-boson system
(namely $E=0.5000 E_{\text{ho}}$) as input for
the zero-range model~III
(this can be viewed as setting the small-$R$ boundary condition;
see Eqs.~(11)-(13) of Ref.~\cite{wernerPRL2006}),
the resulting non-universal
relative
energies for model~III ($2.9019 E_{\text{ho}}$, $5.0587 E_{\text{ho}}$, 
$7.1597 E_{\text{ho}}$, 
and
$9.2351 E_{\text{ho}}$)
agree reasonably
well with those obtained for model~II (see Table~\ref{tab_n3}).
The deviations of around 1~\% are
attributed to the finite 
$r_0$ and $R_0$ values of the two- and three-body potentials.
The above analysis shows that 
model~II provides three-body energies at unitarity that are close
to those for model~III.

\begin{widetext}

\begin{table}
  \begin{center}
  \begin{tabular}{c|c|c|c|c|c|c|l}
    state no.  & $E/ E_{\text{ho}}$ & $\nu$ & $q$ & $C_2 a_{\text{ho}}$ & $C_3 a_{\text{ho}}^2$ & probability & comment \\
    \hline
  1& 0.5000 & 0 & 0 & 35.6 & 0.478 & 0.412 & non-universal \\
  2& 2.9217 & 0 & 1 & 22.4 & 0.334 & 0.426 & non-universal \\
  3& 5.0895 & 0 & 2 & 18.8 & 0.323 & $2 \times 10^{-4}$ & non-universal \\
  4& 5.5187 & 1 & 0 & 21.6 & $9 \times 10^{-6}$ & 0.096 & universal  \\
  5& 7.2005 & 0 & 3 & 17.3 & 0.326 & $1 \times 10^{-4}$  & non-universal \\
  6& 7.5258 & 1 & 1 & 20.5 & $3 \times 10^{-5}$ & 0.027 &  universal \\
  7& 7.8452 & 2 & 0 & 7.15 & $2 \times 10^{-7}$ & 0.003 &  universal \\
  8& 9.2874 & 0 & 4 & 15.4 & 0.336 & $5 \times 10^{-5}$ &  non-universal
  \end{tabular}
  \caption{Three-boson properties at unitarity for
    model~II with $V_0=97,700 E_{\text{ho}}$.
    Column~1 lists the state number.
    Column~2 reports
    the relative energy.
    Columns~3 and 4 report the $\nu$ and $q$
    quantum numbers.
    Columns~5 and 6 report the two- and three-body contacts 
$C_2$ and $C_3$, respectively.
    Column~7 lists the square of the overlap for
    the states at unitarity with an eigen state
of model~II with $v_0=0$ and $V_0=97,700 E_{\text{ho}}$.
    The occupation probabilities for the states
listed add up to
    $0.964$.
  Column~8 indicates whether the state is universal or non-universal.
The energies are estimated to be converged to $0.0010E_{\text{ho}}$
or better.
}
\label{tab_n3}
  \end{center}
  \end{table}

\end{widetext}

In addition to the energies, we calculate the two- and three-body
contacts $C_2$ and $C_3$~\cite{smith2014},
\begin{eqnarray}
\label{eq_c2}
C_2 = - \frac{8 \pi m}{\hbar^2}  \frac{\partial E}{\partial(a_s^{-1})}
\end{eqnarray}
and
\begin{eqnarray}
\label{eq_c3}
C_3 = -\frac{m \kappa_{\text{fs}}}{2 \hbar^2 } 
\frac{\partial E}{\partial \kappa_{\text{fs}}}.
\end{eqnarray} 
In practice, 
the derivative on the right hand side of Eq.~(\ref{eq_c2})
is calculated by changing $v_0$ while keeping $V_0$ constant
and using finite differencing.
Changing $v_0$ translates into
a change of the free-space two-body scattering length $a_s$.
The derivative on the right hand side of Eq.~(\ref{eq_c3}),
in turn,
is calculated by changing $V_0$ while keeping $v_0$ constant.
Changing $V_0$ translates into
a change of the lowest relative free-space three-body 
energy $E_{\text{fs}}$ and thus of the corresponding
binding momentum $\kappa_{\text{fs}}$.
Columns~5 and 6 of Table~\ref{tab_n3} show the
two- and three-body contacts for the harmonically trapped
three-boson system at unitarity.
The universal states are characterized by an extremely
small $C_3$ ($C_3$ should be zero for model~III) since
the likelihood of finding three bosons in close vicinity
to each other is zero
for this class of states. 
The normalization of the three-body contact $C_3$ 
in Eq.~(\ref{eq_c3}) is chosen such that
$C_3=(\kappa^*_{\text{fs}})^2$ for a free-space Efimov trimer
at unitarity
described by model~III. 
For model~II with $V_0=97,700 E_{\text{ho}}$,
the three-body contact for the lowest 
free-space Efimov trimer is, converted to
trap units (using $r_0=0.025a_{\text{ho}}$),
equal to $0.437 (a_{\text{ho}})^{-2}$.
Table~\ref{tab_n3} shows that
the addition of the external confinement leads to a
slight increase
of $C_3$ for the $(\nu,q)=(0,0)$ state
[$C_3=0.478 (a_{\text{ho}})^{-2}$]. 
This makes sense 
intuitively since 
the trap forces the free-space Efimov trimer
into a ``smaller space''.
The two-body contact $C_2$ is roughly comparable
for the universal and non-universal states.

Next, we look at how the energy levels 
evolve as we go from
infinite
to finite to vanishingly small positive
$s$-wave scattering lengths (see 
Refs.~\cite{blume2002,jonsell2002} for early studies).
The energy of all the states shown in Fig.~\ref{fig_energy_n3}
decreases with increasing $a_{\text{ho}}/a_s$,
as the two-body potential becomes deeper.
The state with relative energy $5.5187 E_{\text{ho}}$
at unitarity, i.e., the lowest universal state,
goes through a sequence of 
fairly narrow avoided crossings and
approaches, on the scale of Fig.~\ref{fig_energy_n3}, 
the relative energy for the hard core interaction
model around $a_{\text{ho}}/a_s=2$
(solid line in Fig.~\ref{fig_energy_n3}).
If we prepare the three-boson system
in the 
non-interacting state (which is universal) and then adiabatically
increase the $s$-wave scattering length,
the system will---neglecting the avoided crossings
(they could be jumped across)---end up in the lowest 
universal state at unitarity.
In this sense, we identify this state as 
representing the ``BEC state'', all
the way to unitarity.

Alternatively, we may consider
preparing the three-boson system
in the
lowest
non-interacting state and then instantaneously jumping the
scattering length to infinity.
In this scenario, the occupation probability
of each of the states at unitarity is given by the square of the 
overlap between the initial state and the respective state at
unitarity (see also Ref.~\cite{sykes2014}). 
Table~\ref{tab_n3} shows the occupation probabilities
for an initial state that does not feel a two-body
potential ($v_0$ is set to zero)
but does feel the three-body Gaussian potential.
The relative energy of this state is very close
to that of the non-interacting
state ($3.0004  E_{\text{ho}}$ compared to $3 E_{\text{ho}}$).
The considered quench leads predominantly
to the occupation of non-universal states.
Roughly speaking, the states at unitarity that have an energy
that is comparable to the energy of the initial
state have the largest occupation probabilities.
This observation is consistent with the discussion
presented in Ref.~\cite{sykes2014}. While the occupation
probabilities of the non-universal states  depend on the
three-body parameter, those of the universal states are 
to a very good approximation independent of the
height $V_0$ of the repulsive three-body
Gaussian potential. 

\begin{figure}[t]
  \vspace*{0.1in}
\centering
\includegraphics[angle=0,width=0.4\textwidth]{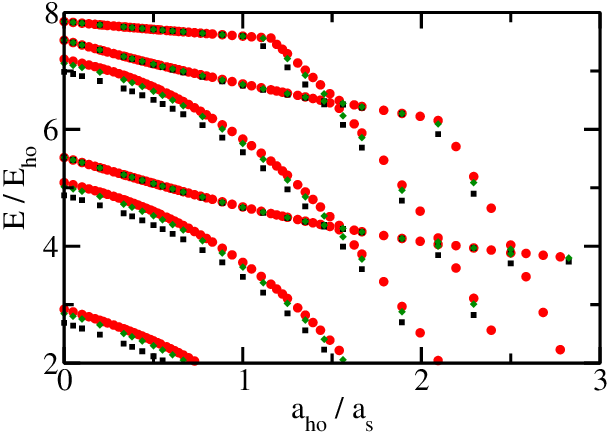}
\vspace*{.5in}
\caption{(Color online)
  Dependence of the energy spectrum for three identical
  harmonically
trapped 
bosons on
the height $V_0$ of the three-body Gaussian potential.
The circles, diamonds, and squares show 
the relative energy as a function of $a_{\text{ho}}/a_s$
for $V_0=97,700 E_{\text{ho}}$ (same as in Fig.~\protect\ref{fig_energy_n3}),
$V_0=40,000 E_{\text{ho}}$, and $V_0=10,000 E_{\text{ho}}$,
respectively.
Only states with $L^{\Pi}=0^+$ symmetry are considered.
}\label{fig_energy_n3_2}
\end{figure}

To investigate the dependence of the three-body energies
on the height $V_0$
of the repulsive three-body Gaussian potential explicitly
over the entire scattering length regime, we 
calculate the three-body spectrum for two other $V_0$ values,
namely $V_0=10,000 E_{\text{ho}}$ and $40,000 E_{\text{ho}}$.
The resulting energies are shown 
by squares and diamonds in Fig.~\ref{fig_energy_n3_2}
together with the energies for $V_0=97,700E_{\text{ho}}$
(circles).
Away from the avoided crossings, the states follow one of
two behaviors:
The energies are to a good approximation independent of
$V_0$ or the energies 
display an appreciable dependence on $V_0$.
This provides a means to classify the states,
away from the avoided crossings, as universal and 
non-universal not only at unitarity but also
for finite scattering lengths.

\section{$N=4$ system}
\label{sec_n4}

\begin{figure}[t]
  \vspace*{0.5in}
\hspace*{0.1in}
\centering
\includegraphics[angle=0,width=0.35\textwidth]{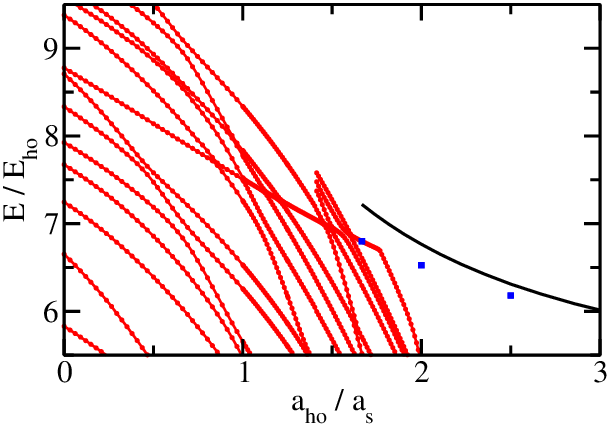}
\hspace*{0.1in}
\vspace*{1.5in}
\caption{(Color online)
Energy spectrum for four harmonically
trapped identical bosons as a function of $a_{\text{ho}}/a_s$.
Only states with $L^{\Pi}=0^+$ symmetry are considered.
The filled circles (neighboring points are connected by
solid lines) show the relative energies for
model~II with $V_0=97,700 E_{\text{ho}}$.
The number of states considered is larger for
$a_{\text{ho}}/a_s \gtrsim 1.4$ than
for  $a_{\text{ho}}/a_s \lesssim 1.4$.
The squares show the relative energies for model~II
with $V_0=97,700 E_{\text{ho}}$
obtained using the ``target state approach''
[these energies are not necessarily upper bounds; moreover, the
errorbar (not shown)
for $a_{\text{ho}}/a_s=2.5$ is comparatively large].
For comparison, the solid line shows the relative ground state
energy for two-body hard core interactions (model~I).
}\label{fig_energy_n4}
\end{figure}

\begin{figure}[t]
  \vspace*{0.1in}
\centering
\includegraphics[angle=0,width=0.4\textwidth]{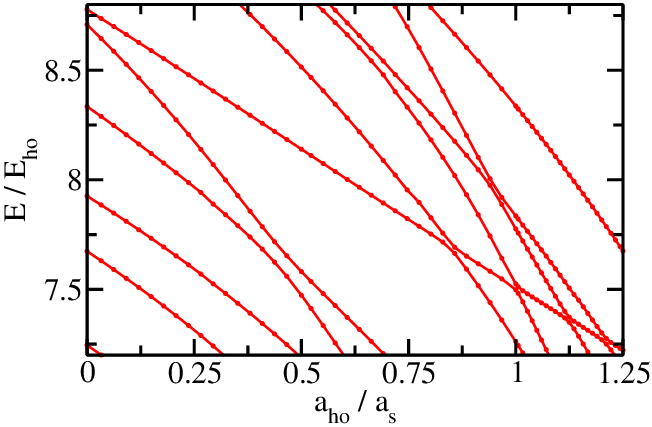}
\vspace*{.5in}
\caption{(Color online)
  Blow-up of a portion of the energy spectrum
  shown in Fig.~\protect\ref{fig_energy_n4}.
    It can be seen that the avoided crossings are quite narrow.
}\label{fig_energy_n4_blowup}
\end{figure}

This section summarizes our results for the harmonically trapped
four-boson system.
Circles in Fig.~\ref{fig_energy_n4} show the $L^{\Pi}=0^+$
energy spectrum for model~II
with the height of the repulsive Gaussian potential
set to $V_0=97,700 E_{\text{ho}}$
(this is the same $V_0$ as that used
in Fig.~\ref{fig_energy_n3})
as a function of $a_{\text{ho}}/a_s$.
At first glance, the energy spectrum, with its many avoided crossings
(a blow-up is shown in Fig.~\ref{fig_energy_n4_blowup}),
looks quite
``messy''.
The four-boson Hamiltonian supports many energy levels that are positive 
for infinite scattering length and go, sometimes after
passing through multiple avoided crossings, to negative energies
as $a_{\text{ho}}/a_s$ increases, i.e., as the two-body potential
becomes deeper. 
Based on the $N=3$ analysis presented in the previous section,
we expect that these rapidly falling four-body levels
correspond to non-universal states.
The four-boson Hamiltonian also supports a few energy levels that
change less with increasing $a_{\text{ho}}/a_s$ and remain positive 
for the largest $a_{\text{ho}}/a_s$ (smallest $s$-wave scattering lengths)
considered.
The lowest of these energy levels plays, as will be 
shown below, a role similar to
the lowest universal state of the three-boson system.

We start our discussion by analyzing the
unitary regime.
The energies of the lowest 26 states
at unitarity are listed in column~2 of Table~\ref{tab_n4}.
To assign the (approximate) hyperangular
and hyperradial quantum numbers $\nu$ and 
$q$
(see columns~3 and 4 of Table~\ref{tab_n4}),
we followed---inspired by
the discussion presented in Sec.~\ref{sec_n3}---a multi-step process.
Note that the quantum number $\nu$ represents an index
that counts the different presumed hyperspherical potential curves.
In a more complete analysis one might hope to
identify a set of quantum numbers for the multiple
hyperangular degrees of freedom.

\begin{widetext}

\begin{table}
  \begin{center}
  \begin{tabular}{c|c|c|c|c|c| c|l}
  state no.  & $E / E_{\text{ho}}$ & $\nu$ & $q$ & $C_2 a_{\text{ho}}$ & $C_3 a_{\text{ho}}^3$ & probability & comment \\
    \hline
    1  & $-0.0622$ & 0 & 0 & 55.0 & 1.35& $0.224$ & strongly non-universal \\
    2  & 2.7139 & 0 & 1 & 39.6 & 0.895 & $0.455$ & strongly non-universal \\
    3  & 4.5646 & 1 & 0 & 54.2 & 0.318 & $0.008$ & strongly non-universal \\
    4  & 5.0460 & 0 & 2 & 35.9 & 0.785 & $0.016$ & strongly non-universal \\
    5  & 5.8297 & 2 & 0 & 33.7 & 0.384 & $0.160$ & strongly non-universal \\
    6  & 6.6516 & 1 & 1 & 51.1 & 0.295 & $0.003$ & strongly non-universal \\
    7  & 7.2457 & 0 & 3 & 34.6 & 0.744 & $1 \times 10^{-4}$ & strongly non-universal \\
    8  & 7.6741 & 3 & 0 & 33.3 & 0.200 & $0.002$ & strongly non-universal \\
    9  & 7.9257 & 2 & 1 & 31.1 & 0.367 & $0.015$ & strongly non-universal \\
    10  & 8.3345 & 4 & 0 & 34.8 & $2 \times 10^{-4}$ & $0.017$ &  universal \\
    11  & 8.7082 & 1 & 2 & 48.6 & 0.278 & $4 \times 10^{-4}$ & strongly non-universal \\
    12  & 8.7773 & 5 & 0 & 31.3 & 0.095 & $0.020$ &  quasi-universal\\
    13  & 9.3789 & 0 & 4 & 35.0 & 0.694 & $5 \times 10^{-4}$ & strongly non-universal \\
    14  & 9.5758 & 6 & 0 & 30.0 & 0.206 & $0.003$ & strongly non-universal \\
    15  & 9.7113 & 3 & 1 & 31.7 & 0.214 & $7 \times 10^{-6}$ & strongly non-universal \\
    16  & 9.9968 & 2 & 2 & 29.3 & 0.370 & $0.004$ & strongly non-universal \\
    17  & 10.3381 & 4 & 1& 33.9 & $3 \times 10^{-4}$ & $0.007$ &  universal \\
    18  & 10.4857 & 7 & 0 & 19.0 & 0.061 & $3 \times 10^{-4}$ &  quasi-universal \\
    19  & 10.5462 & 8 & 0 & 16.0 & 0.045 & $0.001$ &  quasi-universal \\
    20  & 10.7616 & 1 & 3 & 45.6 & 0.258 & $7 \times 10^{-5}$ & strongly non-universal \\
    21  & 10.7895 & 5 & 1 & 30.3 & 0.099 & $0.009$ &  quasi-universal \\
    22  & 10.8732 & 9 & 0 & 22.9 & 0.106 & $0.001$ &  quasi-universal \\
    23  & 11.2272 & 10 & 0 & 16.2 & 0.004 & $0.001$ & universal \\
    24  & 11.4668 & 0 & 5 & 35.8 & 0.562 & $9 \times 10^{-4}$ & strongly non-universal \\
    25  & 11.6047 & 6 & 1 & 27.9 & 0.321 & $0.001$ & strongly non-universal \\
    26  & 11.7560  & 3 & 2 & 30.1 & 0.235 & $1 \times 10^{-4}$ & strongly non-universal \\
  \end{tabular}
  \caption{Four-boson properties at unitarity for model~II
with $V_0=97,700 E_{\text{ho}}$.
    Column~1 lists the state number.
    Column~2 reports
    the relative energy $E$.
    Columns~3 and 4 report the $\nu$ and $q$
    quantum numbers.
Columns~5 and 6 list the two- and three-body contacts $C_2$ and $C_3$,
respectively.
    Column~7 lists the square of the overlap for
    the states at unitarity with an eigen state
of model~II with $v_0=0$ and $V_0=97,700 E_{\text{ho}}$.
    The occupation probabilities for the states
listed add up to
    $0.952$.
Column~8 indicates whether the state is universal,
quasi-universal, or strongly non-universal.}
\label{tab_n4}
  \end{center}
  \end{table}

\end{widetext}

First, we vary $V_0$ while keeping the two-body potential fixed.
Energy levels that do not
move when $V_0$ is changed over a reasonable range are identified
as universal; for these states, the three-body contact
$C_3$ for model~II with $V_0 =97,700 E_{\text{ho}}$
is equal to $0.004 (a_{\text{ho}})^{-2}$ or
smaller (see column~5 of Table~\ref{tab_n4}).
Energy levels that do 
move when $V_0$ is changed are identified
as non-universal; for these states,
the three-body contact
$C_3$ for model~II with $V_0 =97,700 E_{\text{ho}}$
lies between $0.045 (a_{\text{ho}})^{-2}$ and $1.35 (a_{\text{ho}})^{-2}$.
Interestingly, the majority of the low-lying states is 
non-universal.

In the absence of the external confinement, the $(\nu,q)=(0,0)$
state is tied to the lowest free-space Efimov trimer.
For $V_0=97,700 E_{\text{ho}}$, the energy ratio between the 
$N=4$ and $N=3$ free-space energies is $4.59$, which
is quite close to the zero-range value of $4.6108$~\cite{deltuva2010}.
The three-body contact of the trapped four-boson 
states in the $\nu=0$ family
is comparatively large [larger than $0.562 (a_{\text{ho}})^{-2}$ 
for the states listed in Table~\ref{tab_n4}].
The three-body contact
for the $(\nu,q)=(0,0)$ state is
$1.35 (a_{\text{ho}})^{-2}$, which is close to
the three-body contact for the lowest free-space
tetramer [the three-body contact of the lowest free-space
tetramer for $V_0=97,700 E_{\text{ho}}$ is equal to $1.37 (a_{\text{ho}})^{-2}$].

The three-body contacts of the other non-universal 
states fall, roughly, into two categories:
either
$C_3$ is around $0.3 (a_{\text{ho}})^{-2}$ 
[this is similar to the three-body contacts of the 
non-universal $N=3$ states [excluding the $(\nu,q)=(0,0)$
state)]; or $C_3$  is less than about $0.1 (a_{\text{ho}})^{-2}$.
We refer to states with $C_3$ less than about 
$0.1 (a_{\text{ho}})^{-2}$ as ``quasi-universal''
(these states display a comparatively weak dependence on $\kappa_{\text{fs}}$)
and to states with $C_3$ greater than about
$0.1 (a_{\text{ho}})^{-2}$ as ``strongly non-universal''
(these states 
display a comparatively strong dependence on $\kappa_{\text{fs}}$);
see column~8 of Table~\ref{tab_n4}.
While this classification scheme is somewhat arbitrary, 
we employ it since it provides a means 
to categorize the sensitivity of the trapped $N=4$
states on the three-body parameter or, equivalently,
the three-body potential.
Within each $\nu$ family, the three- and two-body
contacts decrease or remain (roughly)
the same with increasing $q$;  
the small increase in selected cases might be due to numerical inaccuracies or
might indicate small irregularities on top of the overall pattern.

As a second step in the classification of states,
we consider the spacings of the energies.
The energy spacings
between consecutive
universal states that live in the same effective
hyperradial potential curve should be equal to $2 E_{\text{ho}}$.
Table~\ref{tab_n4} shows that the spacing between
the energies labeled by
$(\nu,q)=(4,0)$ and $(4,1)$ is very close to $2E_{\text{ho}}$.
The small deviation of $0.004 E_{\text{ho}}$ 
may be due to numerical inaccuracies or due
to the use of the two-body Gaussian potential with finite $r_0$
instead of the two-body zero-range Fermi-Huang pseudopotential.

\begin{figure}[t]
  \vspace*{0.1in}
\centering
\includegraphics[angle=0,width=0.41\textwidth]{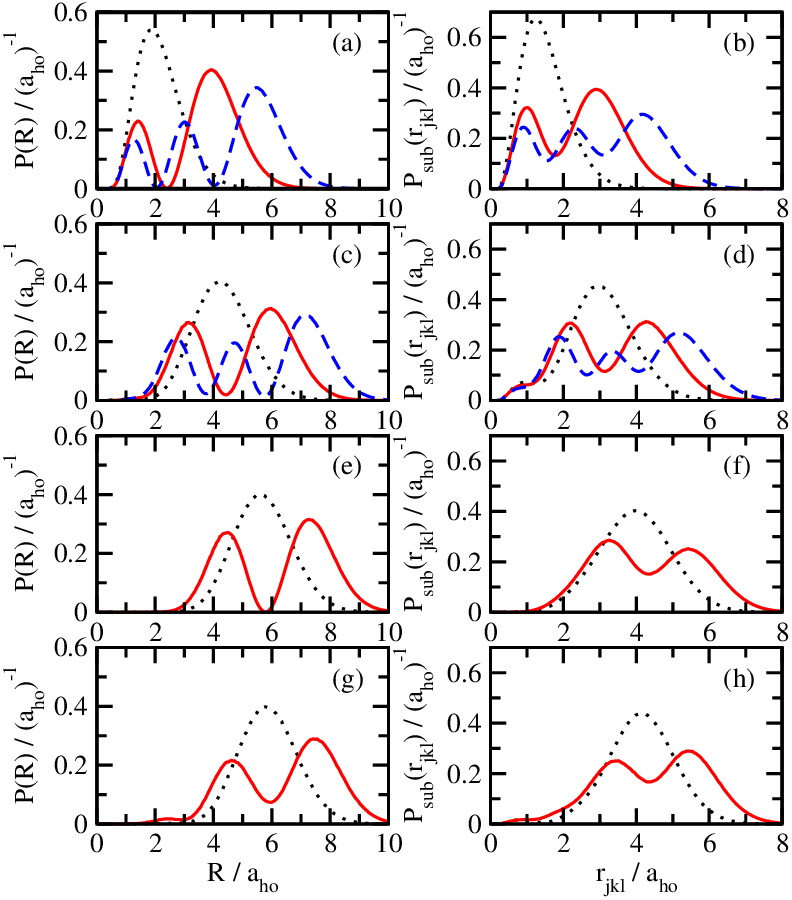}
\vspace*{.5in}
\caption{(Color online)
Structural properties of the harmonically trapped
four-boson system at unitary for interaction model~II with
$V_0=97,700E_{\text{ho}}$.
The hyperradial density $P(R)$ (left column)
and sub-hyperradial density $P_{\text{sub}}(r_{jkl})$ (right column)
are shown for different
$\nu$ families
(the first, second, third, and fourth row
are for $\nu=0$, $\nu=1$, $\nu=4$, and $\nu=5$,
respectively).
The density of the lowest state ($q=0$) in a given
family is shown by a dotted line,
that of the second-lowest state ($q=1$) in a given
family by a solid line, and
that of the third-lowest state ($q=2$) in a given
family by a dashed line.
}\label{fig_hyperradial_n4}
\end{figure}

In addition, we analyze the hyperradial density
$P(R)$, which is normalized such that $\int_0^{\infty} P(R) dR=1$.
The value of the hyperradial density
$P(R)$ tells one the likelihood to find the
four-boson system at the hyperradius $R$.
Figure~\ref{fig_hyperradial_n4}(e) 
shows the hyperradial densities
for the states labeled by
$(\nu,q)=(4,0)$ and $(4,1)$. The hyperradial density for the  $q=1$
states goes to zero at about
$R =5.8 a_{\text{ho}}$, indicating that the corresponding eigen state
can be written, at least to a very good approximation, as a 
product state [see Eq.~(\ref{eq_wavefct_unit_universal})]
even though we are using finite-range interactions. 
The finite ranges
$r_0$ and $R_0$ of our two- and three-body potentials
could, in principle, introduce a small violation of the
separability. For the analysis in this paper, this effect is, however, 
negligible.
This nearly complete factorization further supports the
idea of these $\nu=4$ states being universal,
and clarifies the assignment of the quantum numbers $q=0$ and $1$.


For the non-universal states,
the assignment of the $(\nu,q)$ quantum numbers is
not quite as straightforward.
A key difference between the $N=3$ and $N=4$ systems is that
the non-universal states of the $N=4$ system do not
necessarily separate into a hyperradial piece and a hyperangular
piece (see Sec.~\ref{sec_background}).
One consequence is that the $(\nu,q)$
quantum numbers assigned to the non-universal
$N=4$ states are approximate
and not exact quantum numbers.
A closely related  consequence
is that the hyperradial densities of the non-universal
$N=4$ states do not necessarily vanish for states
with $q>0$. The position of the hyperradial nodes
can depend non-trivially on the hyperangles, implying that
$P(R)$
shows
``washed out'' zeroes.
Figures~\ref{fig_hyperradial_n4}(c)
and \ref{fig_hyperradial_n4}(g) show
examples of this for the
$\nu=1$ and $5$ families, respectively.
For these families, the hyperradial densities for $q>0$
show minima that have finite and not
vanishing amplitude.
Figures~\ref{fig_hyperradial_n4}(a) ($\nu=0$ family),
\ref{fig_hyperradial_n4}(c) ($\nu=1$ family),
\ref{fig_hyperradial_n4}(e) ($\nu=4$ family),
and
\ref{fig_hyperradial_n4}(g) ($\nu=5$ family)
also show that
the hyperradial densities of
states that live, approximately, 
in the same effective hyperradial potential curve 
(labeled by $\nu$) exhibit
similar behavior at small $R$.
This observation lends further support for our 
assignment of the quantum numbers.

The energy level spacings $E_{\nu,q+1}-E_{\nu,q}$
between neighboring non-universal states is
expected to
change (roughly) monotonically from being larger than
$2 E_{\text{ho}}$ for $q=0$ to approximately $2 E_{\text{ho}}$ for large
$q$.
Inspection of Table~\ref{tab_n4} shows that our assignment of the
quantum numbers is consistent with this expectation.

To shed further light on the structural properties of the
trapped four-boson system at unitarity, we take a closer
look at Fig.~\ref{fig_hyperradial_n4}.
Selected aspects of the hyperradial densities $P(R)$ (left column)
were already discussed above. Comparing the
hyperradial densities for $\nu=0$, 1, 4, and 5
[Figs.~\ref{fig_hyperradial_n4}(a),
\ref{fig_hyperradial_n4}(c),
\ref{fig_hyperradial_n4}(e),
and
\ref{fig_hyperradial_n4}(g)],
it can be seen that the $P(R)$
for $\nu=0$ (the $q=0$ state approaches the
lowest four-boson state that is linked to the free-space Efimov trimer
when the trap is removed) has a non-vanishing amplitude at
much smaller $R$ than the states with higher $\nu$.
Moreover, the $\nu=4$ states (these are universal states)
have essentially vanishing amplitude at $R \lesssim 2 a_{\text{ho}}$.
The $(\nu,q)=(5,0)$ state, which is 
quasi-universal and identified below as the BEC state,
displays a small ``bump''
at small $R$,
which we interpret as a signature of the weak dependence on the three-body
potential.
Figures~\ref{fig_hyperradial_n4}(b), \ref{fig_hyperradial_n4}(d),
\ref{fig_hyperradial_n4}(f), and \ref{fig_hyperradial_n4}(h)
show 
the sub-hyperradial density
$P_{\text{sub}}(r_{jkl})$, which corresponds to the 
probability of finding three of the four bosons at a particular
sub-hyperradius $r_{jkl}$.
The strongly non-universal $\nu=0$ and $\nu=1$ states
[Figs.~\ref{fig_hyperradial_n4}(b) and \ref{fig_hyperradial_n4}(d)]
display an appreciable amplitude in the $r_{jkl} \lesssim a_{\text{ho}}$
region.
By comparion, the amplitude of the quasi-universal
$\nu=5$ states [Fig.~\ref{fig_hyperradial_n4}(h)]
is, in the same region, notably smaller
and that
of the universal $\nu=4$ states [Fig.~\ref{fig_hyperradial_n4}(f)]
is essentially zero.

As an aside, we note that the two-body contact 
$C_2$ at unitarity for model~II with
$V_0=97,700 E_{\text{ho}}$ (see column~5 of Table~\ref{tab_n4})
varies by less than a factor of
$4$. This is in contrast to the
three-body contact, which 
varies over roughly
four orders of magnitude.
Generally speaking, the two-body contact
for the trapped $N=4$ system is larger than that for the
trapped $N=3$ system. This makes sense intuitively since the 
four-body system contains twice as many pairs
as the three-body system (six pairs compared to three pairs).
While the two-body contact $C_2$ is obtained by looking
at the variation of the energy with $(a_s)^{-1}$,
it also tells one the likelihood of finding two bosons in close proximity
from each other. The fact that the two-body contact
is of comparable magnitude for the universal and non-universal states
is, at least in part, a consequence of the fact that the two-body
boundary condition is enforced for both classes of states and that
Table~\ref{tab_n4} is limited to the low-energy portion of the
$L^{\Pi}=0^+$ spectrum. We expect 
the two-body contact to be notably smaller for
a subset of the high-lying states and for 
states with finite orbital angular momentum.

As discussed above, the four-boson spectrum depends on
the two-body $s$-wave scattering length (or alternatively, the
lowest relative energy of the trapped two-boson system) and,
in our model~II, the
height of the three-body repulsive Gaussian. The latter can, as done
when calculating the three-body contact, be converted to
the free-space $\kappa^*_{\text{fs}}$ (three-body Efimov parameter)
or, alternatively, the lowest energy of the trapped 
three-boson system at unitarity (this assumes the use of a
low-energy Hamiltonian such that the lowest trimer can be described
by Efimov's theory if the trap is removed).
While we believe that our four-boson spectrum has the same
key
characteristics---such as the energy level crossings
and spacings as well as the classification into
universal and non-universal states---as the ones one would 
obtain for the zero-range model~III,
we did not attempt to extrapolate our results to the 
$r_0 \rightarrow 0$ and
$R_0 \rightarrow 0$ limits.
References~\cite{toelle2011,toelle2013} pursued this, finding
the values of $-0.1(2)E_{\text{ho}}$, $2.7(3)E_{\text{ho}}$, and
$4.6(5)E_{\text{ho}}$ for the lowest three relative four-boson 
energies, for the lowest relative two- and three-boson energies
both fixed at $E_{\text{ho}}/2$.
Our three lowest four-boson energies (see Table~\ref{tab_n4};
due to our finite
value of $r_0$,
our two-boson energy is $0.5103E_{\text{ho}}$ and not $E_{\text{ho}}/2$)
lie within the estimated numerical errorbars of Ref.~\cite{toelle2013}.

Having a solid understanding of the energy spectrum at unitarity,
we consider the finite scattering length regime.
We start our discussion in the weakly-interacting regime,
where the hard core Bose gas model should provide a reasonable description
of the ``lowest BEC state''. Here, the term lowest BEC
state refers to the state that is occupied if the system
is initially prepared in a state with $a_s=0$ and
relative energy $9 E_{\text{ho}}/2$ and if the
$s$-wave scattering length is then increased adiabatically.
The $N=4$ energy of the hard core model is shown by
a solid line in Fig.~\ref{fig_energy_n4} for $a_{\text{ho}}/a_s \ge 1.7$.
While the smallest $a_{\text{ho}}/a_s$ (largest $a_s/a_{\text{ho}}$) considered
may already be somewhat outside of the validity regime
of the hard core model, the energies for the larger $a_{\text{ho}}/a_s$
can be used as a reliable guide for where
the BEC state should lie.
It can be seen that one of the four-boson states for model~II
approaches the hard core model energy with increasing $a_{\text{ho}}/a_s$.
This state undergoes several avoided
crossings. If we diabatize these avoided
crossings (we do this by eye), then this state connects with the lowest
quasi-universal
state at unitarity. We identify this
state as the lowest BEC state.
Interestingly, and somewhat surprisingly,
the lowest BEC state does not, according to our interpretation,
connect to a universal state at unitarity but rather to a 
quasi-universal state.
The implication is that, in a many-body BEC at unitarity,
three-body (or higher-body) parameters may be required to describe the
gas.

The lowest state for $v_0=0$ and $V_0=97,700 E_{\text{ho}}$
has an energy of $4.5017 E_{\text{ho}}$, which is very close to the 
energy of $9 E_{\text{ho}}/2$ of the non-interacting system. Taking this
state as our initial state and assuming an instantaneous sweep
to unitarity, we obtain the occupation probabilities listed in 
column~7 of Table~\ref{tab_n4}.
It can be seen that the $(\nu,q)=(0,0)$ and $(0,1)$
states at unitarity 
have the largest occupation probabilities.
The $(\nu,q)=(0,0)$ state was identified
above as merging into the four-body
state tied to the lowest three-body Efimov trimer when the trapping potential
goes to zero.
Using this, the $(\nu,q)=(0,1)$ 
state, which has the largest occupation probability, 
is best characterized as an ``excited four-body
state'' (breathing mode type excitation)
as opposed to a ``trimer plus atom state''.
This observation begs the question whether the $N$-body 
short-time dynamics is
governed predominantly by three-body Efimov physics (as implicitly
implied if the $N$-body dynamics is modeled by a
three-body Hamiltonian with scaled density)
or whether four-body and possibly also higher-body physics
plays a non-negligible role,
at least for certain parameter combinations.

\begin{figure}[t]
\centering
\includegraphics[angle=0,width=0.4\textwidth]{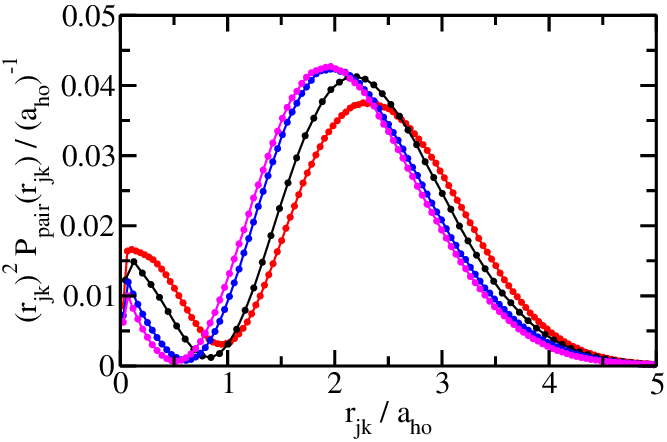}
\vspace*{.5in}
\caption{(Color online)
  Scaled pair distribution function
  $r_{jk}^2 P_{\text{pair}}(r_{jk})$ for the lowest BEC state
  for various $a_s$.
  The calculations are for the interaction model~II with
$V_0=97,700E_{\text{ho}}$.
  The curves from bottom
  to top at $r_{jk}/a_{\text{ho}}=1.5$
  correspond to
  $a_{\text{ho}}/a_s=0$, $1/2$,
  $\approx 1.3500$, and $\approx 1.7012$.
}\label{fig_pair}
\end{figure}

Figure~\ref{fig_pair} shows the
scaled pair distribution function
$r_{jk}^2 P_{\text{pair}}(r_{jk})$,
normalized such that $4 \pi \int P_{\text{pair}}(r_{jk}) r_{jk}^2 d r_{jk}=1$,
of the lowest BEC state for
selected $a_s$.
The scattering length values are chosen
such that the state identified as the lowest
BEC state is isolated (away from avoided crossings).
For the smallest $a_s$ considered, the pair distribution
function
goes to a very good approximation to zero
at $r_{ij}=a_s$.
If we assume a Jastrow-type
variational wave function that consists of a product over two-body
functions,
the observed behavior is consistent with the intuitive
picture that the lowest BEC state is described
by two-body correlation functions that have a node, i.e., that
can be interpreted as excited pairs
(see also Refs.~\cite{ancilotto2015,sze2017}).
As $a_s$ increases, the
state identified as the lowest BEC state 
is characterized by a pair distribution function
that displays a minimum at $r_{jk}$ values smaller than $a_s$.
This can be thought of as a kind of saturation.
Clearly, if a minimum exists it has to be at finite $r_{jk}$ and not
at $r_{jk} \rightarrow \infty$
as
$a_s$ goes to infinity.
Interestingly, the minimum of $r_{jk}^2P(r_{jk})$ does not
go zero but takes on finite values.
This suggests that
higher-body correlations may play a role in the large $a_s$ limit.

\section{Summary and outlook}
\label{sec_conclusion}

This paper presented a first comprehensive
study of the harmonically trapped four-boson system
interacting through two-body short-range interactions with
positive $s$-wave scattering length $a_s$, including infinitely
large $a_s$.
The two-body interactions were parametrized through a
purely attractive two-body Gaussian,
which allows---in free space---for the formation
of many deeply-bound molecular states.
To eliminate deep-lying molecular states and
to work in a regime where the free-space
four-body energies at unitarity are tied
uniquely (i.e., through universal numbers)
to the free-space Efimov trimers at unitarity, a
purely repulsive three-body Gaussian potential was added.
The interaction parameters were adjusted
such that the
size $L_{\text{fs}}$ of the
lowest free-space Efimov trimer at unitarity is just a bit larger
than the characteristic length $a_{\text{ho}}$ of the external
spherically symmetric harmonic
trapping potential. Correspondingly, the size of the lowest
free-space tretramer at unitarity is quite close to $a_{\text{ho}}$.
Since the ``internal'' characteristic length scales
(sizes of the free-space trimer and tetramer)
are comparable to the ``external''
characteristic length scale
(harmonic oscillator length $a_{\text{ho}}$),
the chosen parameter
combinations are expected to yield a rich energy spectrum.
Indeed, the low-energy four-boson energy spectrum
for the employed model interaction displays a
maze of energy levels and crossings.

Among the many energy levels, we identified---diabatizing, by eye,
some of the avoided crossings---one energy level as the lowest
gas-like BEC state.
While a strict definition of this state was not provided,
our working definition was as follows: Assuming one prepares the
four-boson system in the non-interacting eigenstate and one then
increases the $s$-wave scattering length adiabatically,
jumping across narrow avoided crossings quickly,
one should follow an energy level whose energy increases
monotonically and whose properties depend at most weakly
on the three-body interaction employed.
While we quantified the three-body contact at unitarity, we did
not re-calculate the entire four-boson spectrum for a second or
third parametrization of the three-body potential,
using the same $\kappa^*_{\text{fs}}$, i.e., fixing the
``internal'' scale to the same value.
Moreover, this work did not re-calculate the four-boson
energy spectrum for other $\kappa^*_{\text{fs}}$ values.
Studying the dependencies on the parametrization of the
interactions and systematically varying $\kappa^*_{\text{fs}}$ are left
for future work.
Nevertheless, with the energy spectrum at hand, we were able
to extract some information of the lowest gas-like BEC state.
In particular, the pair distribution function in the strongly-interacting
regime acquires a minimum at $r \approx a_{\text{ho}}$ but
does not go to zero. This finding may provide guidance for
constructing improved variational descriptions
of the strongly-interacting Bose gas.

The paper also presented the first comprehensive analysis of
the energies of the trapped four-boson system at unitarity.
The energy levels were assigned approximate quantum numbers and
classified as universal (vanishing three-body contact $C_3$)
and non-universal (finite three-body contact $C_3$).
Moreover, depending on the value of $C_3$, the non-universal states
were further categorized as ``quasi-universal''
and ``strongly non-universal''.
The assignment and classification scheme
of the four-boson system were corroborated by
analyzing structural properties, namely the hyperradial density
$P(R)$ and the sub-hyperradial density
$P_{\text{sub}}(R)$.
Our work suggests that the contacts $C_2$ and $C_3$ can be thought of
as analogs of
quantum numbers in that they provide a classification scheme of the
states.

The present work suggests several future research directions.
It would be interesting to compare the present energy spectrum
with that for other interaction models as well as other
$\kappa^*_{\text{fs}}$.
It would also be interesting to use the four-boson spectrum
presented here as a starting point for simulating time ramps,
possibly including both three- and four-body loss coefficients.
The presented results can be used as benchmarks
with which to test approximate variational schemes,
collective coordinate approaches, or trial wave functions employed
in quantum Monte Carlo studies.

\section{Acknowledgement}
\label{acknowledgement}
DB gratefully
acknowledges support by the National Science Foundation through
grant numbers
PHY-1509892 and PHY-1745142
and discussions with Hans-Werner Hammer.
MWCS and JLB acknowledge support from the 
JILA Physics Frontier Center, grant number PHY-17345142.

\end{document}